\documentclass[12pt]{article}
\usepackage{graphicx}

\newcommand\pubdate{\today}

\def\Title#1{\begin{center} {\Large #1 } \end{center}}
\def\Author#1{\begin{center}{ \sc #1} \end{center}}
\def\Address#1{\begin{center}{ \it #1} \end{center}}

\def\dd{\rm d}
\usepackage{setspace} 

\newcommand\pubblock{\rightline{\begin{tabular}{l}  \\ 
         \pubdate  \end{tabular}}}
\newenvironment{Abstract}{\begin{quotation}  }{\end{quotation}}
\newenvironment{Presented}{\begin{quotation} \begin{center} 
             PRESENTED AT\end{center}\bigskip 
      \begin{center}\begin{large}}{\end{large}\end{center} \end{quotation}}

\begin{document}
\begin{titlepage}
 \pubblock
\vfill
\Title{Neutrino deep-inelastic scattering cross sections \\ from 100 GeV to 1000 EeV}
\vfill
\Author{Daniel R. Stump, on behalf of the CTEQ-TEA Collaboration}
\Address{Michigan State University}
\vfill
\begin{Abstract}
The CTEQ-TEA Collaboration has calculated neutrino-nucleon cross sections, based on CTEQ18 NNLO parton distribution functions.\cite{NuNCT18}
\end{Abstract}
\vfill
\begin{Presented}
DIS2023: XXX International Workshop on Deep-Inelastic Scattering and
Related Subjects, \\
Michigan State University, USA, 27-31 March 2023 \\
\includegraphics[width=9cm]{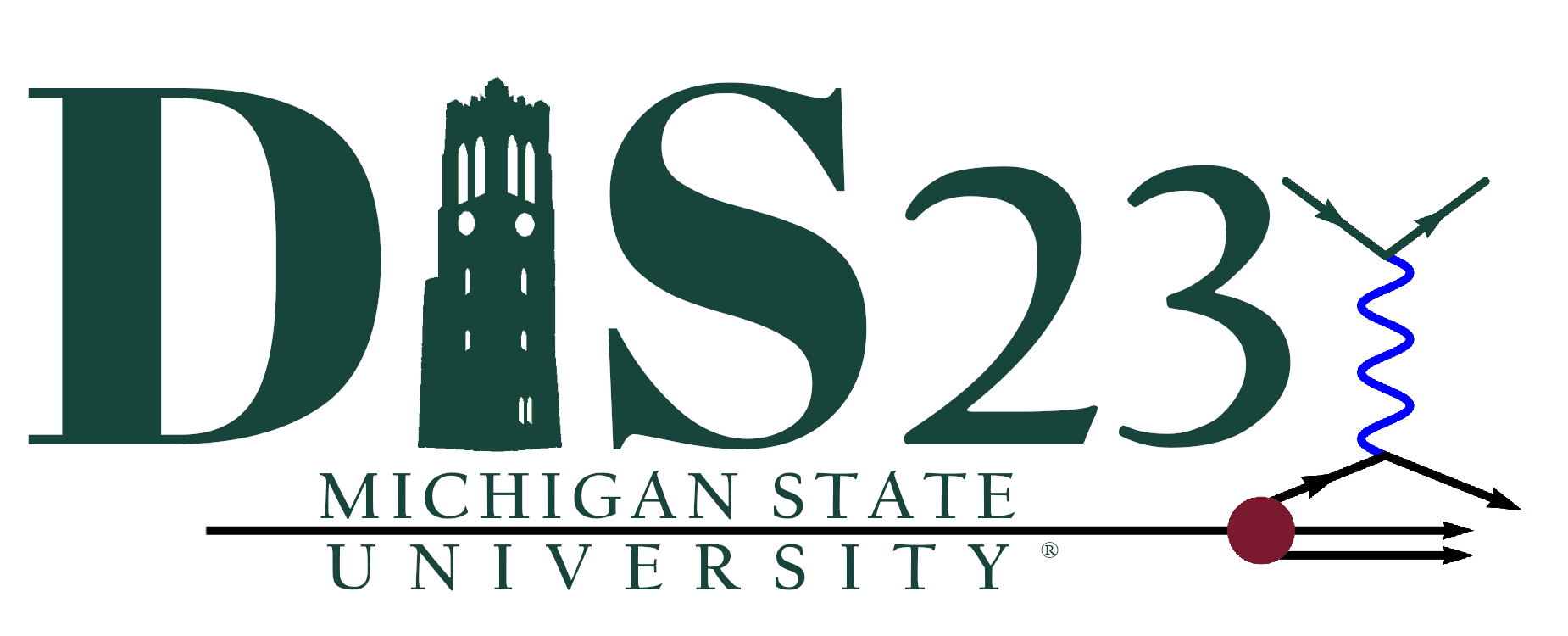}
\end{Presented}
\vfill
\end{titlepage}


\singlespacing

A classic challenge in weak-interaction physics is to measure the cross sections for neutrino interactions.
For example, proton accelerators have been used to create neutrino beams; then charged-current $\nu$ interactions  are observed when neutrinos strike a dense target.
Let $E_{\nu}$ denote the neutrino energy in the rest frame of the detector, and let $\sigma_{\nu}$ denote cross sections (collectively) for $\nu$ or $\overline{\nu}$ hitting the targets.
The energies $E_{\nu}$ from accelerator experiments range from 1 to 350 GeV; the highest $E_{\nu}$ come from the Tevatron, e.g., observed at the NuTeV detector.

Today, neutrino cross sections at significantly higher energies can be measured using neutrinos of  astrophysical origins.
In particular, the IceCube Collaboration has published measurements for 
$10\, {\rm TeV} \leq E_{\nu} \leq 10^{4}\, {\rm TeV}$.\cite{IceCube:2017aar, IceCube:2020rnc}
The measurements rely on theory predictions for such energies. They can be compared to standard-model predictions to search for new physics.
Also, the analysis of IceCube events provides information on the flux of astrophysical neutrinos as a function of $E_{\nu}$.
For these reasons, it is important to have
state-of-the-art predictions of neutrino cross sections $\sigma_{\nu}$ , with, importantly, their theoretical uncertainties $\delta\sigma_{\nu}$.

The IceCube analyses require accurate predictions of $\sigma_{\nu}(E_{\nu})$. In current publications, the analyses are based
on the ``CSMS Model''\cite{Cooper-Sarkar:2011jtt},
in which the total cross section $\sigma_{\nu}(E_{\nu})$
is calculated for deep-inelastic scattering of neutrinos from free nucleons, using HERAPDF parton distribution functions (PDFs). 

The purpose of this contribution to DIS2023 is to show the ``CT18 Model'' predictions, based on the CTEQ18 global analysis of QCD.\cite{NuNCT18,Hou:2019efy}.
Figure 1a shows the Feynman diagram for $\nu$-nucleon DIS. The differential cross section may be written as 
\begin{equation}\label{eq:diffXS}
\frac{\dd^2\sigma^{\nu(\bar{\nu})}}{{\dd}{x}\,{\dd}{Q^2}}=
\frac{G_F^2}{4\pi x(1+Q^2/M_{W,Z}^2)^2}
\left[Y_{+}(y)\,F_2-y^2\,F_L \pm Y_{-}(y)\,xF_3\right],
\end{equation}
where the structure functions $F_{i}(x,Q^2)$
are expressed in terms of PDFs.

Collectively there are eight neutrino-nucleon cross sections, denoted $\sigma_{\nu}$, for 2 projectiles ($\nu$ and $\overline{\nu}$) $\times$ 2 targets (p and n) $\times$ 2 interactions (CC and NC).
Below we will show ``isoscalar nucleon cross sections'': $\sigma_{I} \equiv \frac{1}{2} (\sigma_{p}+\sigma_{n})$,
relevant to a target with equal numbers of free protons and neutrons.
We have also calculated $\sigma_\nu$ for nuclear targets, e.g., O-16, which differs from $8\,{p}+8\,{n}$ by nuclear effects;
the calculations require nuclear PDFs.\cite{NuNCT18}

The CSMS Model has been used consistently by the IceCube Collaboration for their neutrino cross section analyses.
The CSMS Model uses
HERAPDFs $f(x,Q)$.\cite{Cooper-Sarkar:2011jtt}
Data from HERA experiments provide important information at small $x$.
The HERAPDF QCD global analysis is based on
the next-to-leading-order (NLO) QCD perturbation theory.
The HERAPDFs do not include data from LHC experiments.

Our CT18 Model for $\sigma_\nu$\ \cite{NuNCT18}
is based on the CTEQ18 global analysis
of QCD.\cite{Hou:2019efy}
It differs from the CSMS Model in several ways.
Most importantly, {\it (i)} we use the NNLO order of QCD perturbation theory; and {\it (ii)} we include relevant LHC data in the global analysis. 

Judgements that are made in the uncertainty
analysis, are different for CT18 and CSMS.
Any parton theory must provide predictions {\it with uncertainties},
i.e., $\sigma_{\nu} \pm \delta\sigma_{\nu}$,
as a function of $E_{\nu}$.
Both the CT18 and CSMS models use similar Hessian analyses of the PDF uncertainties;
but the two collaborations make different judgements of the net resulting uncertainty on predictions.

Figure 1b shows our CT18 predictions for
$\frac{1}{2}(\sigma^{W}_{\nu}
+\sigma^{W}_{\overline{\nu}})_{\rm isoscalar}$,
along with IceCube measurements.\cite{IceCube:2020rnc}

\begin{figure}
    \centering
\includegraphics[width=0.40\textwidth]{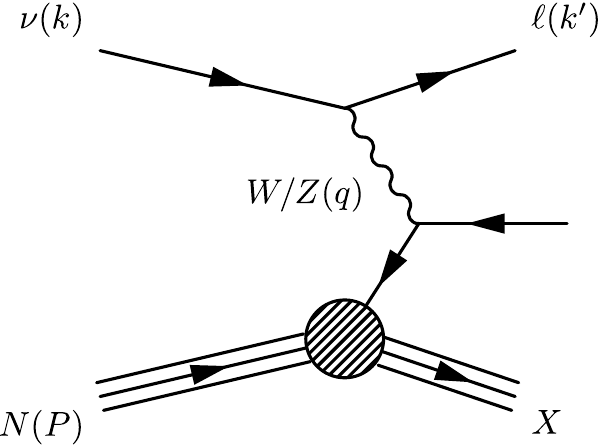} \hfill
\includegraphics[width=0.55\textwidth]{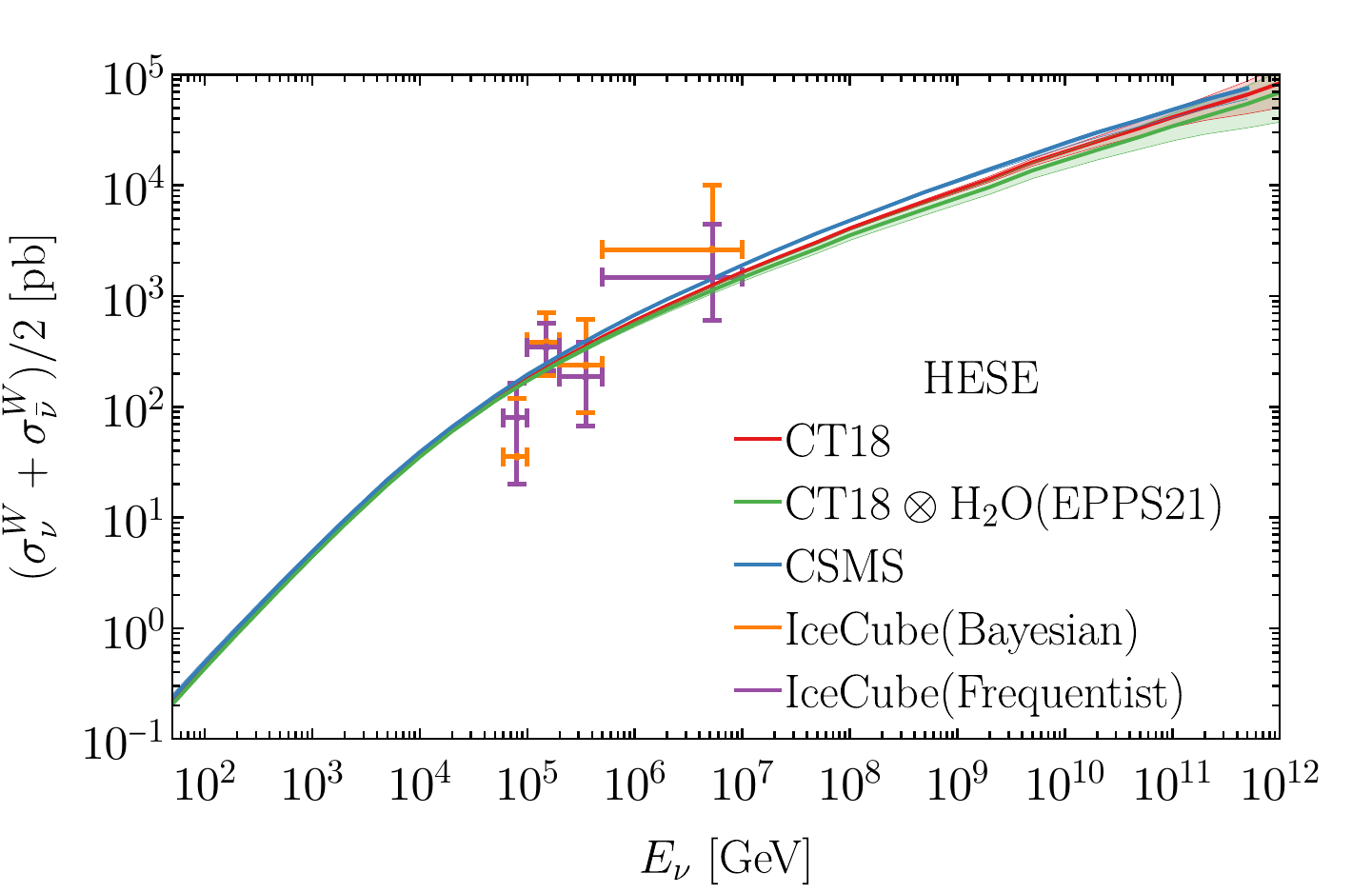} 
\caption{Neutrino-nucleon scattering.
Left: Feynman diagram for deep-inelastic scattering;
Right: Total cross sections from the CT18NNLO calculations,
and IceCube data.}
\label{fig:DRSONE}
\end{figure}

\bigskip

Figure 2 illustrates our CTEQ18  results,
for $\sigma_{\nu}$ versus $E_{\nu}$.
Four cross sections are shown---both neutrino and antineutrino projectiles, and both charged-current and neutral-current interactions---for isoscalar targets of free nucleons, i.e., $\frac{1}{2} \left[ \sigma_p+\sigma_n \right]$.
$E_{\nu}$ is the neutrino energy in the nucleon rest frame.
The center of mass energy,
$\sqrt{s} = \sqrt{2\,m\,E_{\nu}}$,
ranges from 10 GeV to $10^{6}$ GeV.
At small energies we have $\sigma_{\nu} \propto\ E_{\nu}$, corresponding to the approximate 4-fermion coupling $G_{F}$, consistent with accelerator experiments. At higher energies the linear dependence decreases, consistent with the Froissart bound,
due to the finite W and Z masses.
This behavior is seen in IceCube $\sigma_{\nu}$ measurements.

The right panel of Fig.\,{2} shows our
PDF uncertainty $\delta\sigma$ versus $E_{\nu}$,
as a percentage of the central prediction
$\langle\sigma\rangle$.
The uncertainty has been taken to be symmetric,
i.e., $\sigma = \langle{\sigma}\rangle \pm \delta{\sigma}$.
For small $E_{\nu}$ the uncertainty is a few percent of
$\langle\sigma\rangle$,
as expected for an NNLO global analysis.
For large $E_{\nu}$, the uncertainty increases,
becoming more than 50 percent of
$\langle\sigma\rangle$
for the largest $E_{\nu}$ considered.
The reason for this large uncertainty is that 
for high $E_{\nu}$, $\ \sigma_{\nu}$ depends strongly on PDFs at small $x$, i.e., values of $x$ far below
the range of data used in the global analysis.
The PDFs are not well constrained there,
so $\delta\sigma$ is large.

Detailed evaluations of the predictions and uncertainties of $\sigma_{\nu}$ are explained in Ref.\cite{NuNCT18}.
Figure 2 illustrates our final results.
Tables are provided in Ref.\cite{NuNCT18}. 

\begin{figure}
    \centering
\includegraphics[width=0.48\textwidth]{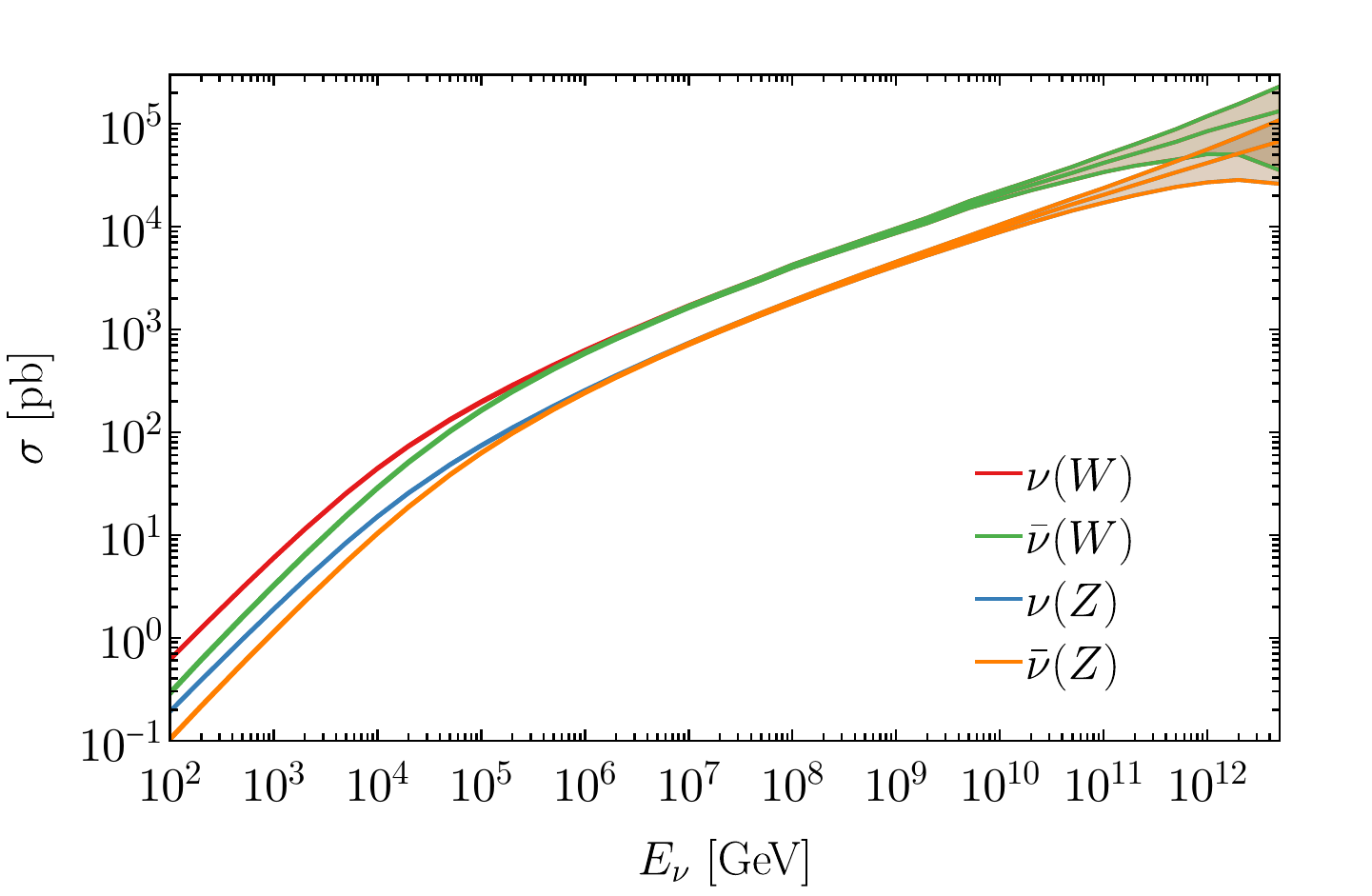} \hfill
\includegraphics[width=0.48\textwidth]{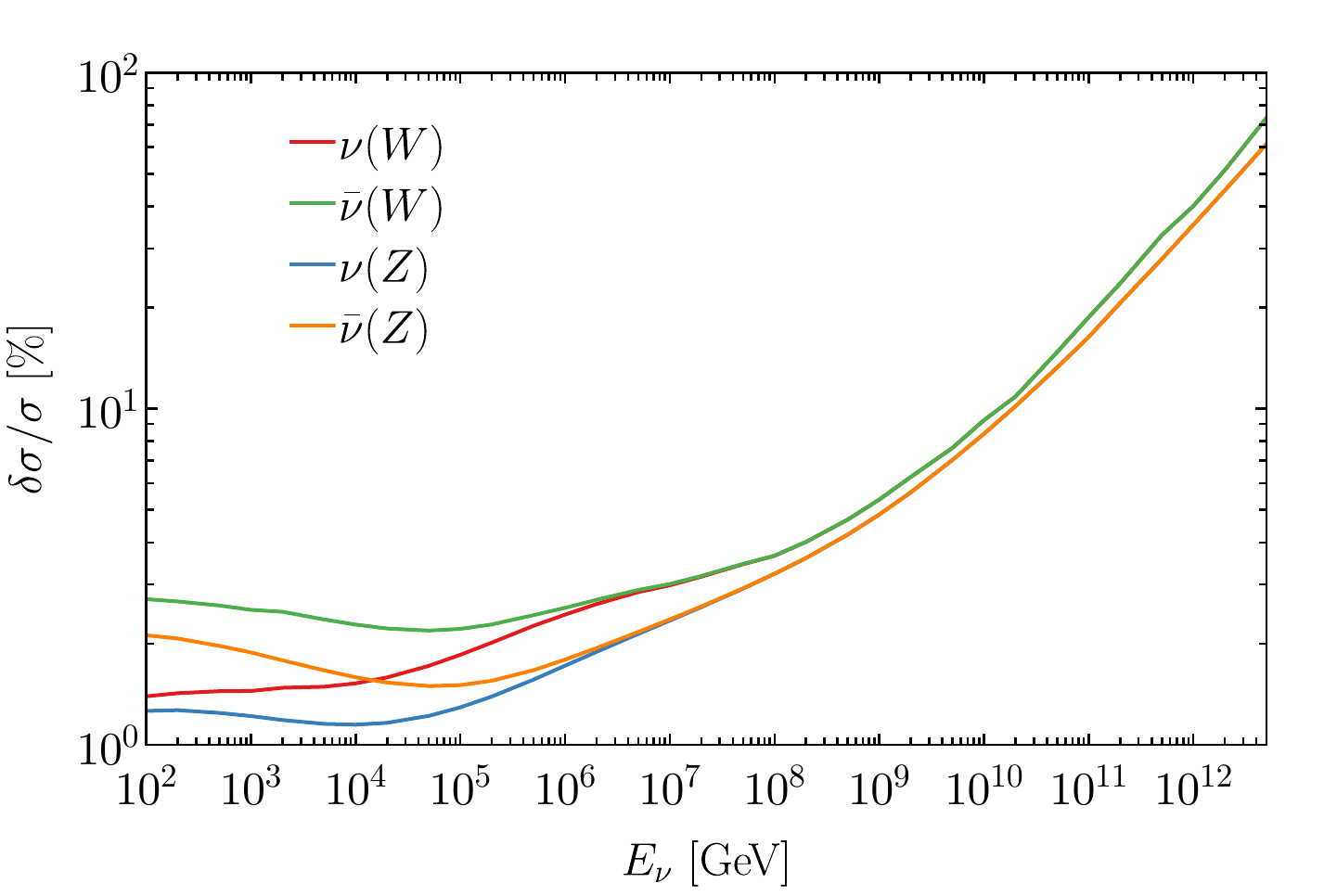} 
\caption{Final results for the CT18 Model
of $\sigma_{\nu}(E_{\nu})$.
Left: Four isoscalar cross sections, including central predictions and uncertaunty bands.
Right: Uncertainties $\delta{\sigma}$ as percentages of the central predictions $\langle\sigma\rangle$.}
\label{fig:DRSTWO}
\end{figure}


\bigskip
\noindent
{\bf Concerning details of the calculations...}

\noindent
{\it  $\bullet\ $ Small-x partons.}
The differential cross section
${\dd}^{2}\sigma/({\dd}x\,{\dd}Q^{2})$
is important in the global analysis of QCD.
Now we want the total cross section,
\begin{equation}
\sigma_{\rm total}(E_{\nu})
= \int_{0}^{2 m E_{\nu}} dQ^{2}\ \int_{0}^{1}\ dx
\ \frac{{\dd}^{2}\sigma}{{\dd}x\, {\dd}Q^{2}}(x,Q^{2}).
\end{equation}
However, PDFs are only published for limited kinematic ranges, $Q > Q_{\rm min}$ and $x > x_{\rm min}$.
For example, the CT18NNLO PDFs are only published for $x > 10^{-6}$.
What functions $f(x,Q)$ should we integrate down to $x=0$?
Also, there may be gluon saturation at small-$x$.
Small-$x$ is important for $\sigma_{\nu}$ at high $E_{\nu}$.
(Cf.\ the talk about small-$x$ PDFs by Keping Xie at the DIS2023 Conference.)

\noindent
{\it  $\bullet\ $ NNLO QCD perturbation theory.}
The CTEQ18 global analysis of QCD provides NNLO PDFs,
and a complete set of uncertainties based on eigenvectors of the Hessian matrix.\cite{Hou:2019efy} 
However, within the NNLO approximation
there are alternative treatments for parton masses;
we use the S-ACOT-$\chi$ formulas.
Also, we tested some known ${\rm N^{3}LO}$ corrections,
and verified that they are very small.

\noindent
{\it  $\bullet\ $ Nuclear effects.}
These are due to strong interactions
inside nuclear targets with mass number $A>1$.
They must be calculated with {\it nuclear PDFs}.
Depending on $A$, the nuclear corrections become comparable or even larger than NNLO-order perturbative QCD corrections.

After studying the options we made specific choices for our final predictions:
CT18NNLO PDFs, six quark flavors, some ${\rm N^{3}LO}$
corrections, and next-to-leading log(x) corrections
at small $x$.\cite{NuNCT18}
Our tables provide
the eight cross sections, with PDF uncertainties.
Also, tables for an O-16 nuclear target are given.

\bigskip

The IceCube Collaboration has consistently used the CSMS model of $\sigma_{\nu}(E_{\nu})$ to analyze the rate of neutrino events at the IceCube Observatory.
So next we compare our CT18 Model to the CSMS results.
Figure 3 (left panel) shows one comparison:
the ratio $\sigma_{\rm CSMS}/\sigma_{\rm CT18}$ for the full range of $E_{\nu}$, for neutrino CC events.
(Antineutrino and NC events are similar.\cite{NuNCT18})
The CSMS Model of $\sigma_{\nu}$ is larger,
by $\sim{5}$ percent at small $E_{\nu}$,
up to $\sim{20}$ percent at large $E_{\nu}$.
The CT18 and CSMS uncertainty bands for the ratio are also shown in Fig.\,3a, and are comparable.
However, the fact that the uncertainty bands do not overlap, implies that there are significant differences between the models.

The differences in Fig.\,3a have two interesting causes.
First,  the difference between NLO and NNLO parton distributions accounts for part of the difference.
We have compared CT18NLO to CT18NNLO,
and find that the NLO PDFs do give larger neutrino cross
sections similar to Fig.\,3a.
Second, the LHC data used in the CTEQ18 global analysis
(not included in HERAPDF) affects the PDFs;
e.g., the gluon distributions
of CTEQ18 and HERAPDF differ in central value and uncertainty.
Details of the differences between CTEQ18 and HERAPDF
PDFs are shown in Ref.\cite{NuNCT18},
to account for the differences
between the predictions of $\sigma_{\nu}$.

Figure 3a also compares the CT18NNLO free nucleon $\sigma_{\nu}$,
to the cross section for an O-16 nucleus target, which we calculated with O-16 PDFs.\cite{NuNCT18, Eskola:2021nhw}.
Here we see that the nuclear cross section is
$\sim{10}$ percent smaller than the CT18 model for
free nucleons;
the difference is {\it a nuclear
shadowing effect} at small x;
also, the nuclear uncertainty is larger.

\begin{figure}
    \centering
\includegraphics[width=0.48\textwidth]{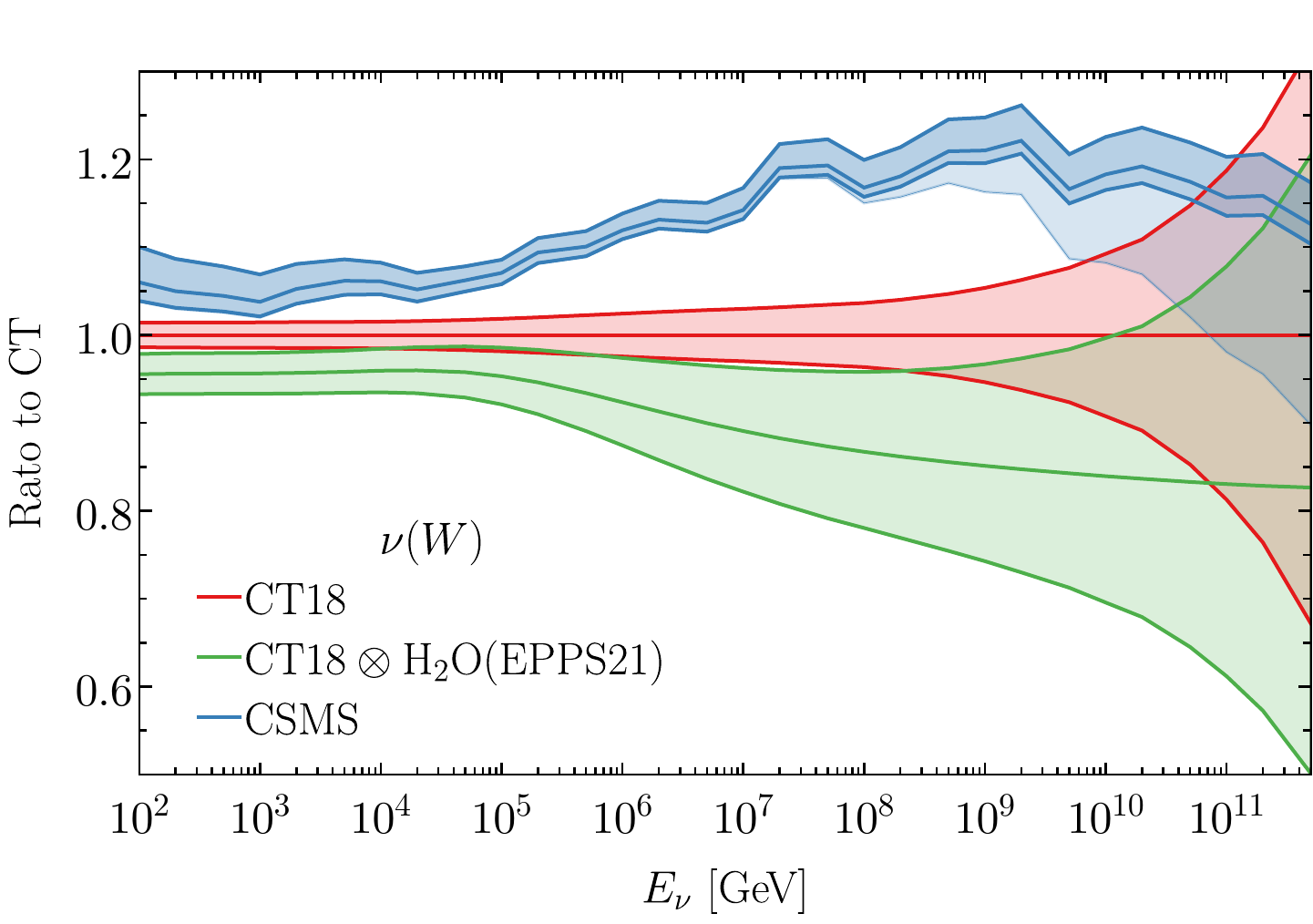} \hfill
\includegraphics[width=0.48\textwidth]{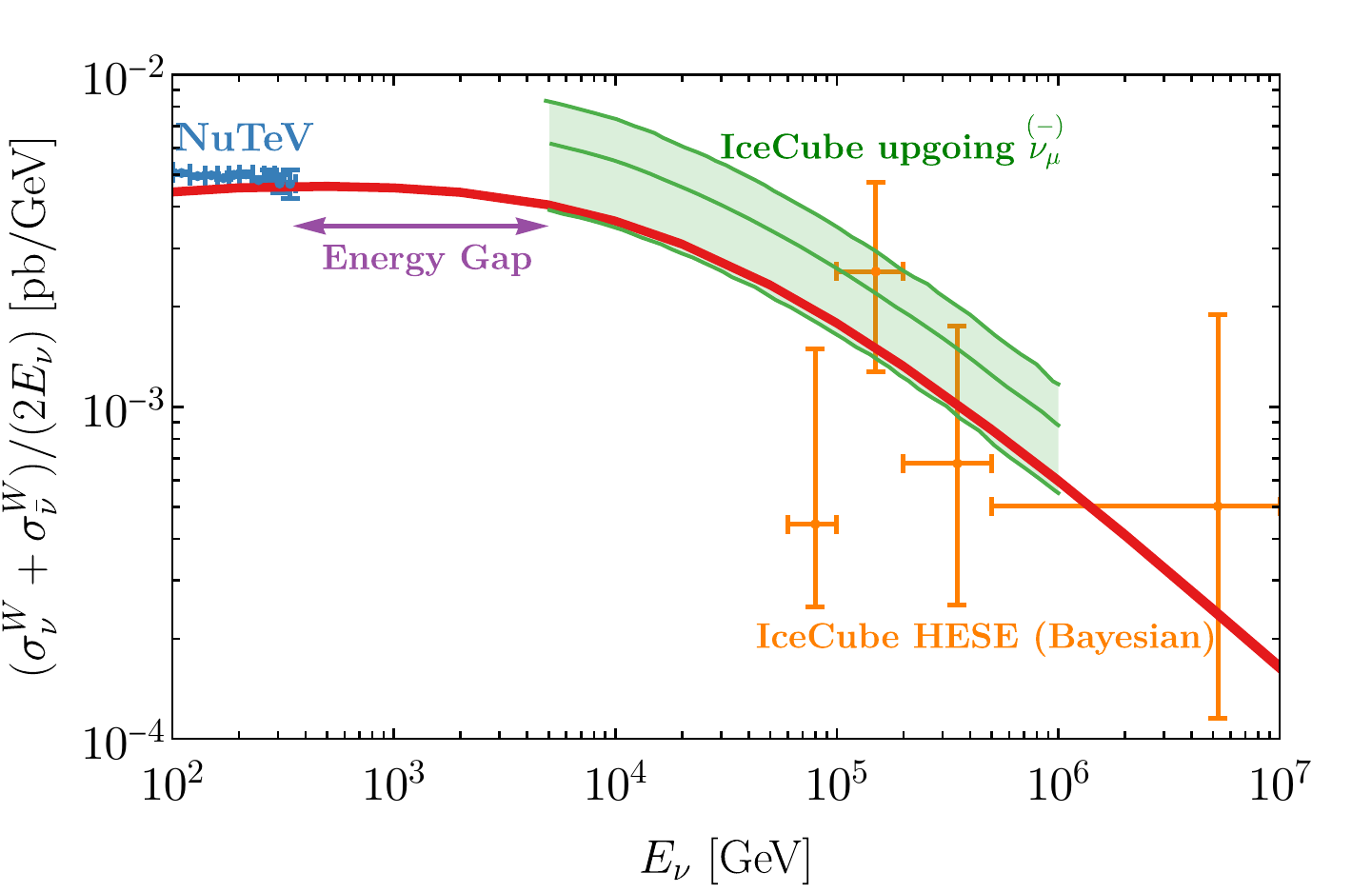} 
\caption{
Left: Comparison of CT18 and CSMS Models for $\sigma(\nu,CC,{\rm isoscalar})$; the graph shows
ratios of $\sigma_{\nu}$
to the central CTEQ18 prediction.
The $\sigma_{\nu}$ for an O-16 target is also shown (EPPS21 nuclear PDFs\cite{Eskola:2021nhw}).
Right: Comparisons of predictions to current and future data.}
\label{fig:proc3}
\end{figure}

\bigskip

The theory predictions of $\sigma_{\nu}(E_{\nu})$ are important for the IceCube Neutrino Observatory. 
IceCube is an ice Cherenkov detector in Antarctica, which observes charged leptons produced by CC interactions of high-energy neutrinos.
During approximately 10 years of operation, it has collected cosmic neutrino events at high $E_{\nu}$.

Both CC and NC cross sections are important for the IceCube analysis.
Only CC events are observed in the detector.
However,  one method to measure the cosmic neutrino flux $\Phi$ is to analyze ``absorption'' of neutrinos by the Earth.\cite{IceCube:2017aar}
The absorption, or scattering,
is due to both CC and NC events.
The full analysis yields an estimate of $\Phi$,
together with a test of the neutrino cross sections
used in the analysis.
Figures 1b and 3b show some IceCube data.
Although the error bars are large,
this is just the beginning. 

The IceCube Collaboration has proposed upgrades of the Observatory, to achieve higher statistics, and to extend the range to higher $E_{\nu}$.

Figure 3 (right) illustrates the past, present, and future of neutrino cross sections.
The graph is $\frac{1}{2}\left(
\sigma^{W}_{\nu}+\sigma^{W}_{\overline{\nu}}\right)/E_{\nu}$
versus $E_{\nu}$.
At low $E_{\nu}$, past accelerator experiments have reached $E_{\nu}$ up to $\sim{350}$ GeV, which is important in the global analysis of QCD and PDFs.
Now IceCube energies extend to
$10\,{\rm TeV} \leq E_{\nu} \leq 10\,{\rm PeV}$,
which can search for new physics and for astrophysical sources of high-energy neutrinos.
There is an interesting energy gap from 350 GeV to 10 TeV, which will be filled in at the LHC Run 3,
using forward detectors, such as FASER $\nu$.


\begin{thebibliography}{99}
\bibitem{NuNCT18}
Keping Xie, Jun Gao, T.\ J.\ Hobbs, Daniel R.\ Stump and C.-P.\ Yuan,
``High-energy neutrino deeply inelastic scattering cross sections from 100 GeV to 1000 Eev'',
(preprint report numbers: ANL-181430, MSUHEP-23-003, PITT-PACC-2302);
arXiv:2303.13607[hep-ph].

\bibitem{Hou:2019efy}
Tie-Jiun Hou, and others (CTEQ-TEA Collaboration),
Phys.\ Rev.\ D 103, 014013 (2021);
arXiv:1912.10053[hep-ph].

\bibitem{IceCube:2017aar}
IceCube Collaboration, M.\,G.\,Aartsen et al,
Nature 551 (2017) 596-600;
arXiv:1711.08119[hep-ex]

\bibitem{IceCube:2020rnc}
R. Abbasi and others (IceCube Collaboration),
Phys.\ Rev.\ D 104, 022001 (2021);
arXiv:2011.03560[hep-ex].

\bibitem{Cooper-Sarkar:2011jtt}
Amanda Cooper-Sarkar, Philipp Mertsch and Subir Sarkar,
JHEP 08 (2011) 042;
arXiv:1106.3723[hep-ph].

\bibitem{Eskola:2021nhw}
Kari J.\ Eskola, Petja Paakkinen, Hannu Paukkunen and Carlos A.\ Salgado,
Eur.\ Phys.\ J.\ C 82, 413 (2022);
arXiv:2112.12462[hep-ph].

\end{thebibliography}
\end{document}